\documentclass[conference]{IEEEtran}
\IEEEoverridecommandlockouts
\usepackage{cite}
\usepackage{amsmath,amssymb,amsfonts}
\usepackage{algorithmic}
\usepackage{graphicx}
\usepackage{textcomp}
\usepackage{xcolor}
\usepackage{import}
\usepackage{url}
\usepackage[hypertexnames=false,bookmarks=false]{hyperref} 
\hypersetup{colorlinks,linkcolor={red},citecolor={blue},urlcolor={black}} 
\usepackage[euler]{textgreek} 

\usepackage{ragged2e}

\ifCLASSOPTIONcompsoc
    \usepackage[caption=false, font=normalsize, labelfont=sf, textfont=sf]{subfig}
\else
\usepackage[caption=false, font=footnotesize]{subfig}
\fi

\def\BibTeX{{\rm B\kern-.05em{\sc i\kern-.025em b}\kern-.08em
    T\kern-.1667em\lower.7ex\hbox{E}\kern-.125emX}}

\begin{document}
\bstctlcite{IEEEexample:BSTcontrol}
\addtolength{\textheight}{7.5mm}

\title{Open-Source Memory Compiler for Automatic RRAM Generation and Verification\\}

\author{
\IEEEauthorblockN{Dimitrios (Dimitris) Antoniadis$^*$, Peilong Feng$^{*\dagger}$, Andrea Mifsud$^{*\dagger}$, Timothy G. Constandinou$^{*\dagger\ddag}$}
\IEEEauthorblockA{$^*$Department of Electrical and Electronic Engineering, Imperial College London, SW7 2BT, UK\\$^\dagger$Centre for Bio-Inspired Technology, Institute of Biomedical Engineering, Imperial College London, SW7 2AZ, UK\\$^\ddag$Care Research \& Technology Centre, UK Dementia Research Institute, UK\\
\{dimitris.antoniadis20, peilong.feng14, a.mifsud, t.constandinou\}@imperial.ac.uk\\
}
}

\maketitle

\begin{abstract}
 The lack of open-source memory compilers in academia typically causes significant delays in research and design implementations. This paper presents an open-source memory compiler that is directly integrated within the Cadence Virtuoso environment using physical verification tools provided by Mentor Graphics (Calibre). It facilitates the entire memory generation process from netlist generation to layout implementation, 
 and physical implementation verification. To the best of our knowledge, this is the first open-source memory compiler that has been developed specifically to automate Resistive Random Access Memory (RRAM) generation. RRAM holds the promise of achieving high speed, high density and non-volatility. 
 A novel RRAM architecture, additionally is proposed, and a number of generated RRAM arrays are evaluated to identify their worst case control line parasitics and worst case settling time across the memristors of their cells.
     The total capacitance of lines SEL, N and P is 5.83\,fF/cell, 3.31\,fF/cell and 2.48\,fF/cell respectively, while the total calculated resistance for SEL is 1.28\,\textOmega/cell and 0.14\,\textOmega/cell for both N and P lines. 

\end{abstract}

\section{Introduction}

Random Access Memories (RAMs) and processing units are critical components of a digital system. Commonly known volatile memories are Static Random Access Memories (SRAMs) and Dynamic Random Access Memories (DRAMs)~\cite{zahoor2020resistive, chen2016review}. SRAMs are of particular interest since they can be used in System-On-Chip (SoC), Application Specific Integrated Circuit (ASIC) and microprocessor designs~\cite{guthaus2016openram, xu2007flexible}. 

Memory design is however a tedious task for an Integrated Circuit (IC) design engineer, and requires plethora of time and multiple design cycles~\cite{guthaus2016openram,xu2007flexible}. Additionally, commercial Process Design Kits (PDKs) that are available to academia or small and medium size enterprises (SMEs), do not provide memory compilers. Access to memory IP is typically licensed through third party vendors, often providing just limited flexibility and reconfigurability~\cite{guthaus2016openram}. Modern applications require customised memories (such as single port, dual port, multi port)~\cite{shah2010fabmem}. Depending on the desired performance, different fabrication technology may be required~\cite{guthaus2016openram}, therefore, memory compilers should be able to create scalable and customisable memories~\cite{guthaus2016openram,xu2007flexible}. Driven by the above mentioned needs, various attempts have been made to create memory compilers~\cite{guthaus2016openram,xu2007flexible,shah2010fabmem,wu201065nm}.

Conventionally, memory compilers mainly deal with SRAMs or other volatile memories~\cite{guthaus2016openram,xu2007flexible,shah2010fabmem,wu201065nm,goldman2014synopsys,clinton20185ghz}. 
These memories can have advantages such as high packing density and high read/write speed. Nevertheless, the volatility of the states of their cells produces the necessity of constantly refreshing their cell values (DRAMs) or restoring them on power-up (SRAMs). As a result the energy consumption is high~\cite{zahoor2020resistive, maheshwari_hybrid_2020}. On the other hand, commercially available non-volatile memories (flash memories), suffer of low speed, high write voltage and low endurance~\cite{zahoor2020resistive}. Both volatile and non-volatile memories face the problem of \textit{the memory wall}, as the write/read speed of memories does not scale at the same rate as the operation frequency of processing units does~\cite{wulf1995hitting,ielmini2018memory,chen2016review}.


It is clear that novel material memories exploiting new memory architectures are necessary to achieve non-volatility, scalability, high speed, high density and low power performance~\cite{zahoor2020resistive,chen2016review,ielmini2018memory,stathopoulos2019electrical, yang2013memristive}. These devices can be characterised as emerging Non-Volatile Memory (NVM) technologies~\cite{zahoor2020resistive,stathopoulos2019electrical}. Such a device is the \textit{memristor}~\cite{yang2013memristive,strukov2008missing, maheshwari_hybrid_2020}. A promising type of memory using memristors and addressing the above mentioned problems is the Resistive RAM (RRAM)~\cite{chen2016review, maheshwari_hybrid_2020, lee2019reram}.

\begin{figure}[!t]

\centerline{\includegraphics[width=0.95\columnwidth]{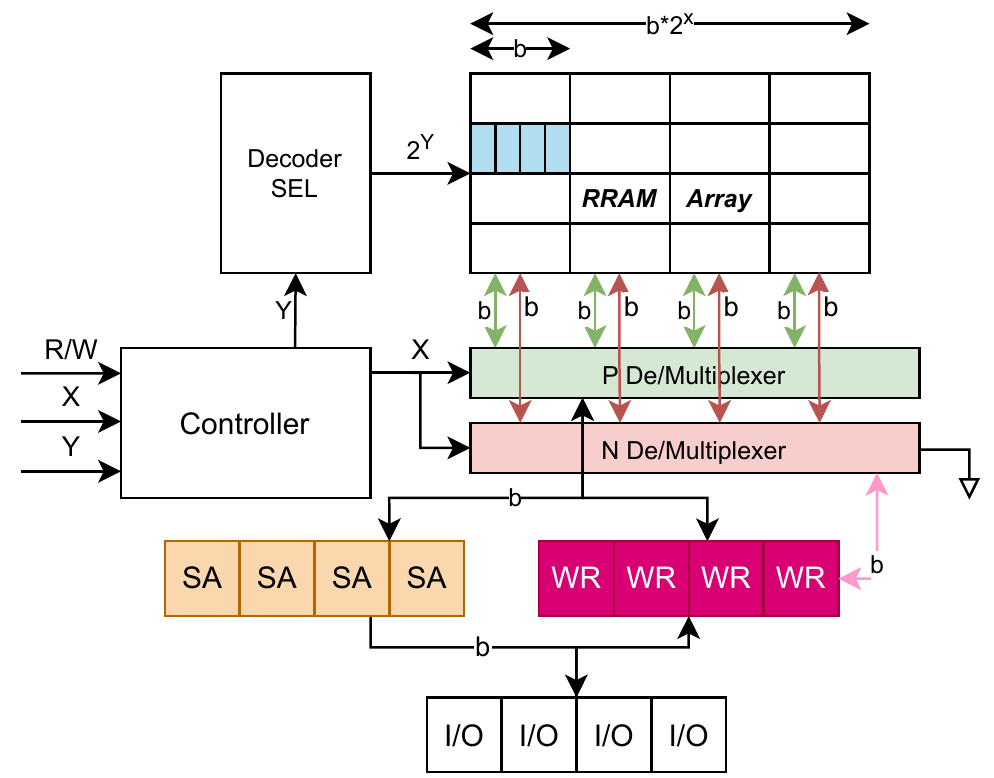}}
\vspace{-3mm}
\caption{Architecture of proposed RRAM: Top Right - RRAM array. The size of the array is $M=2^Y$ rows and $N=b2^X$ columns. The blue coloured cells represent a word of $b$ bits. Left -  A $Y$ bits to $2^Y$ bits Decoder to choose the address (enable horizontal SEL line) of the desired word. Bottom - two $b2^X$ to $b$ De/Multiplexers (enable vertical P, N lines) to provide the necessary connections based on the read/write operation. These are connected to the Sense Amplifier Array and the Write Circuits Array shown in orange and purple colour respectively.}
\label{fig:RRAM_architecture}
\vspace{-6mm}
\end{figure}

Even though, a RRAM compiler has been presented before \cite{lee2019reram}, the authors of this paper acknowledge the need of an open-source RRAM compiler for academia and industry. The compiler should be able to generate a RRAM and its peripheral circuits. Additionally, the compiler should provide automatic layout verification and  memory characterisation. The proposed tool should be independent of fabrication technology and should provide reconfigurability options on the architectures and the circuits of the memory. Such a tool is essential in order to investigate novel properties and capabilities of RRAMs. This is developed as part of the Functional Oxide Reconfigurable Technologies (FORTE) programme~\footnote{For latest news on FORTE and resources visit \url{www.forte.ac.uk}}. FORTE's aims are threefold: develop and optimise the manufacturing processes of RRAM devices; create the technology tools and design rules required for integration with CMOS; and showcase potential applications of CMOS/Memristor integrated circuits~\cite{maheshwari_hybrid_2020}.

This paper presents an early version of the first open-source RRAM compiler which automatically creates and verifies the schematic and the layout of a M $\times$ N dimensions RRAM. Additionally, it extracts parasitics and generates a post layout view. Suitable files with parasitic information are generated which can be imported in MATLAB and further information on control lines capacitance and resistance can be obtained. The RRAM compiler has been published on Github:

\begin{center}
    \url{https://github.com/akdimitri/RRAM_COMPILER}
\end{center}

The proposed architecture of the RRAM of the compiler is shown on Fig.~\ref{fig:RRAM_architecture}. The subcircuits are quite similar to an SRAM~\cite{wicht2003current}, however, RRAM architecture requires additional peripheral circuitry to control the RRAM cells which operate on very strict voltage conditions as  explained in Section~\ref{sec:RRAM_cell_architecture}.
Section~\ref{sec:Compiler} describes the operation flow of the proposed compiler. Section~\ref{sec:results} compares the characteristics of the various RRAM macros created by the compiler. Finally, Section~\ref{sec:conlusions} discusses future improvements, challenges and concludes the results.

\section{RRAM Architecture Overview}\label{sec:RRAM_cell_architecture}

\begin{figure}[tb]
\centerline{
\includegraphics[width=0.85\columnwidth]{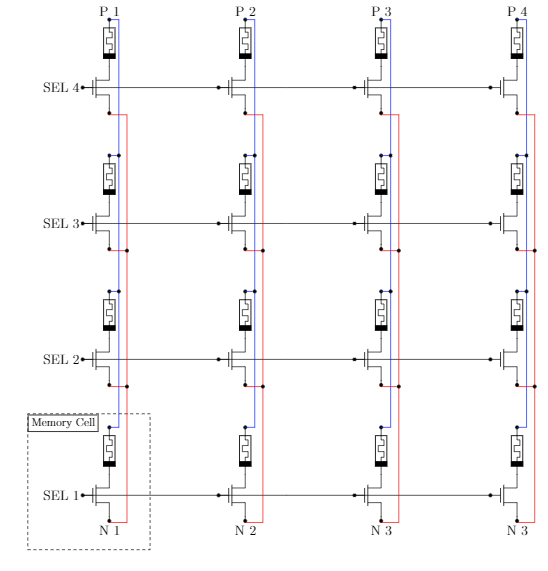}
}
\vspace{-3mm}
\caption{RRAM array $4 \times 4$. RRAM rows share SEL line while RRAM columns share P (blue) and N (red) lines. One memory cell is shown on the lower left part of the image surrounded by dashed line.}
\label{fig:RRAM_array}
\vspace{-5mm}
\end{figure}

The proposed RRAM cell consists of a memristor and an NMOS transistor as shown in Fig.~\ref{fig:RRAM_array}. In order to write a value to the cell, the SEL line has to be set to high voltage and an appropriate voltage difference has to be applied between P and N terminals (depending on memristor characteristics). In order to read the value of the memristor, SEL line has to be set to high voltage and an extremely small voltage difference has to be applied between P and N terminals, ensuring the voltage difference is not large enough to change the resistive state of the memristor. Thus, it is made clear that suitable peripheral circuits have to be designed in order to accurately control these sensitive pins.

\begin{figure}[tb]
\centerline{\includegraphics[width=0.95\linewidth]{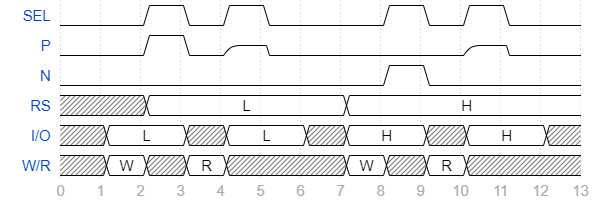}
}
\vspace{-2mm}
\caption{Read and Write operations. Assuming an unknown Resistive State RS on the memristor, at $t=1$ a write signal appears on write/read (W/R) port and Low Resistance ($L$) is received on Input/Output (I/O) port. At $t=3$ the read operation of the cell begins. Note the small voltage applied on P line while reading function of RRAM. At $t=8$, a high voltage pulse is applied on SEL and N line in order to write a High Resistance ($H$) value.} 
\label{fig:RRAM_readwrite}
\vspace{-4mm}
\end{figure}


The RRAM cells can form a RRAM array by sharing horizontal SEL lines and vertical P and N lines as it is shown on Fig.~\ref{fig:RRAM_array}. Based on this topology a M $\times$ N RRAM array can be generated. Given that a word length is $b$ bits, a $2^Y \times b2^X$ RRAM array can be generated, where $M=2^Y$ and $N=b2^X$ with resemblance to the arrays of Figs.~\ref{fig:RRAM_architecture} and \ref{fig:RRAM_array}.

Thus for both read and write operation, a SEL line decoder is needed to select the desired horizontal line and multiplexers to connect the P and N vertical lines of the desired word memory cells to read/write circuits. Since the array has $2^Y$ SEL lines, a word of $Y$ bits is enough to describe the address of the desired SEL line. Therefore a decoder of $Y$ to $2^Y$ is required.

Regarding P and N lines, it is essential to take under consideration the read and write operations. During the read operation, a small voltage (less than 0.5\,V) has to be applied on the desired P lines. Therefore, a multiplexer $b2^X$ to $b$ is necessary to connect the selected memory word to read circuits. A word of $X$ bits is enough to control the multiplexer. Since the voltage difference has to be extremely small, the N pins can be connected to ground (See Fig.~\ref{fig:RRAM_architecture} ground on N Multiplexer). Therefore, a signal can be used (not shown on figure) during read operation to ground all N pins. On the other hand, during write operation a larger voltage value has to be applied across P and N pins, thus in this case both desired P and N lines will be connected to write circuits using $b2^X$ to $b$ multiplexers. A simplified diagram of Read and Write Operations is shown on Fig.~\ref{fig:RRAM_readwrite}. In Fig.~\ref{fig:RRAM_architecture} N multiplexer is connected with pink line with the write circuits as it is needed only for write operation. On the other hand, P multiplexer is connected on both read and write circuits, thus a black line is used on the figure. The write circuits are shown on Fig.~\ref{fig:RRAM_architecture} with WR blocks.

As far as the read operation is concerned, a novel sense amplifier has to be designed. In contrast to conventional SRAM sense amplifiers and other sense amplifiers for NVMs, which apply a minimum voltage of $1/3$\emph{VDD} or greater on the bit line~\cite{wicht2003current,blalock1991subnanosecond,chang2015read}, the sense amplifier of the proposed RRAM has to maintain an extremely small voltage on P line throughout the whole read operation, in order not to change the resistive state of the memristor. Potentially, no precharge circuit is required in contrast to SRAM implementations~\cite{guthaus2016openram}. The sense amplifiers array is shown on Fig.~\ref{fig:RRAM_architecture} with SA blocks. These operations require a control logic circuit that will supervise the whole operation of the RRAM circuits. Besides the abstract system architecture presented on this section, further circuits are required, such as Tri-state input/output ports. These kind of circuits and additional peripheral circuits are essential for the RRAM operation and will be investigated and added at a later stage of the Compiler design.




\begin{figure}[tb]
\centerline{
\includegraphics[width=0.95\columnwidth]{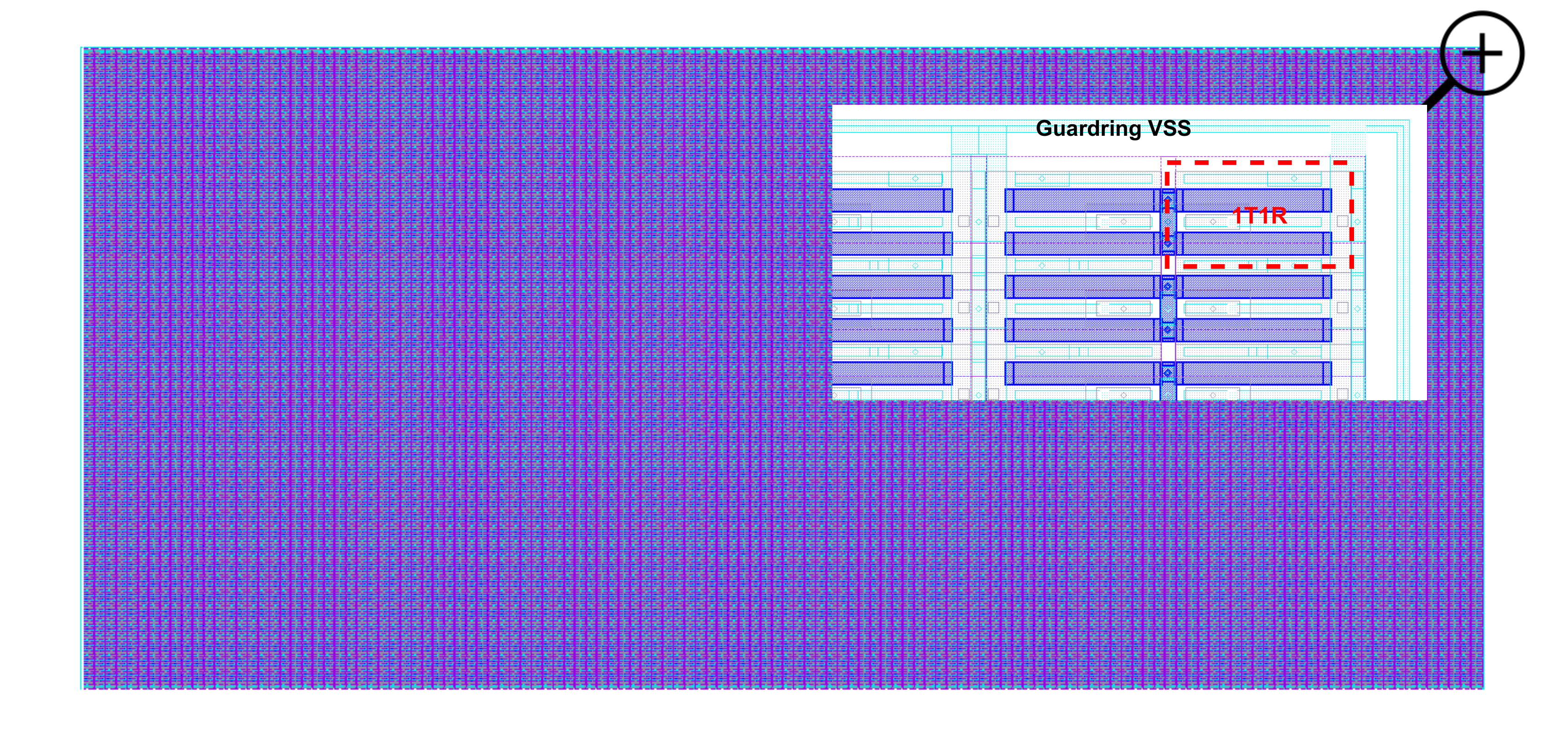}
}
\vspace{-4mm}
\caption{Generated $128 \times 128$ RRAM array  layout. On the top right of the picture, a zoomed area is shown (poly, metal 1, metal 4). It depicts a memory cell (1T1R) and the shared connections.}
\label{fig:RRAM_array_layout}
\vspace{-4mm}
\end{figure}
\section{Compiler Analysis} \label{sec:Compiler}

The proposed compiler was created using \textit{SKILL} language which can be used to customise and extend Cadence Design Environment (IC6.1.8)\cite{skillRef,skilluser}. In this case, multiple SKILL scripts and functions have been created and integrated in a main function SKILL script which performs the automatic memory generation and layout verification. Both the main function and  subfunctions are generic functions, highly customisable and reconfigurable. 

The arguments of the compiler are the $M$ x $N$ dimensions and the name of the library where the generated RRAM should be saved. Additionally, the database representation~\cite{skillRef} of the RRAM cell and the Calibre cell-map file can be provided as arguments or they can be directly hard-coded. 

\begin{figure}[tb]
\centerline{
\includegraphics[width=\columnwidth]{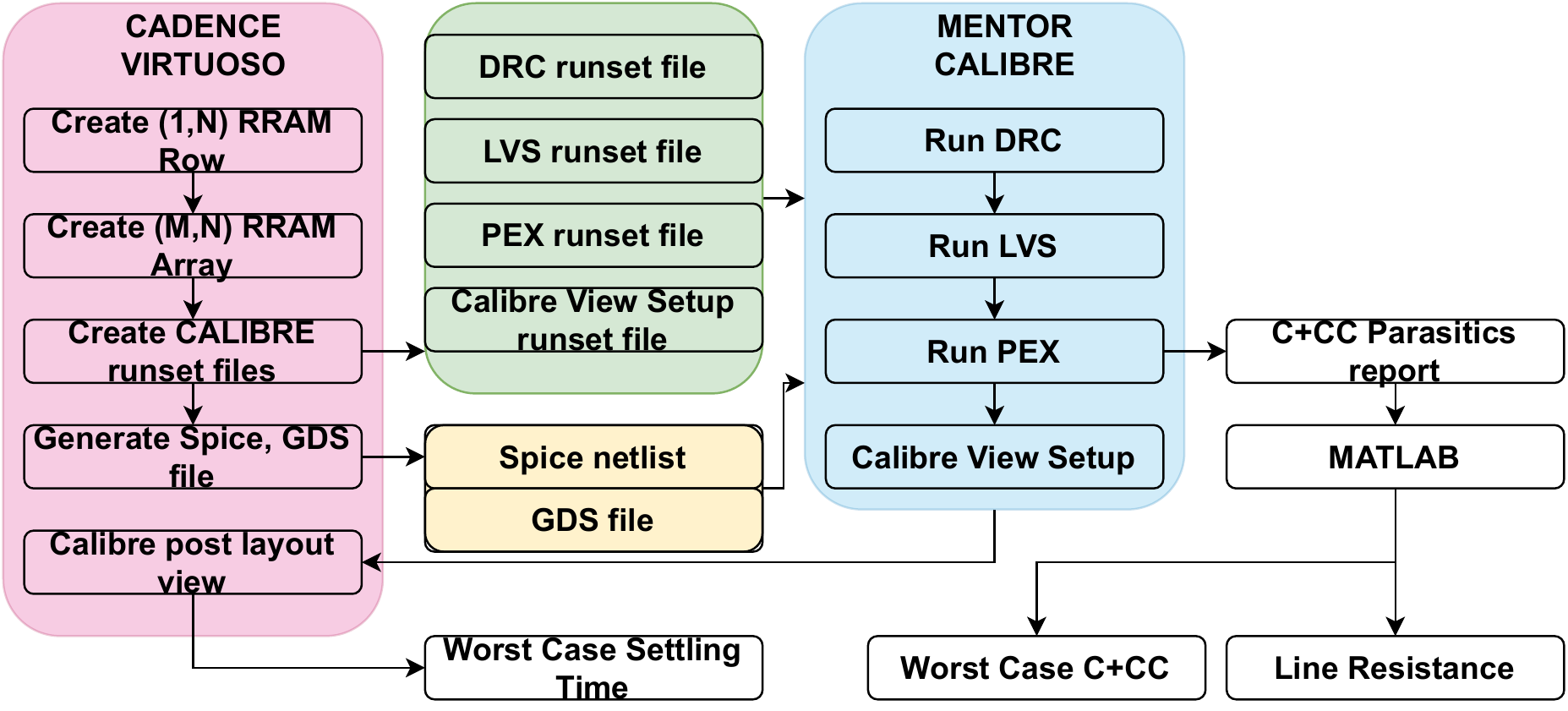}
}
\vspace{-2mm}
\caption{Flowchart of RRAM compiler operations.}
\label{fig:RRAM_flow}
\vspace{-5mm}
\end{figure}

A simplified compilation flow is shown on Fig.~\ref{fig:RRAM_flow}. Initially the compiler creates the schematic, the layout and the symbol of a row of RRAM Cells of size $1 \times N$. Based on the row cellview, the RRAM macro is generated. 

Subsequently, the compiler creates three folders inside the cellview folder for DRC, LVS and PEX respectively. The necessary files (runset files, Spice netlist, GDS) are generated by the compiler in order to invoke Calibre DRC, LVS and PEX in batch mode through the Command Interpreter Window (CIW) of Virtuoso~\cite{calibreuser, calibreinter}. After these operations are finished, the compiler invokes Calibre View Setup in batch mode and creates a Calibre post-layout view of the RRAM. The parasitics output files of the compiler are located inside the cellview folders. 

The layout (metals and poly layers) of a generated $128 \times 128$ RRAM  array is presented on Fig.~\ref{fig:RRAM_array_layout}. The layout size is 294.42\,\textmu m$\times$642.41\,\textmu m.

\section{Results} \label{sec:results}

\begin{figure}[!tb]
\centering
\includegraphics[width=0.9\linewidth]{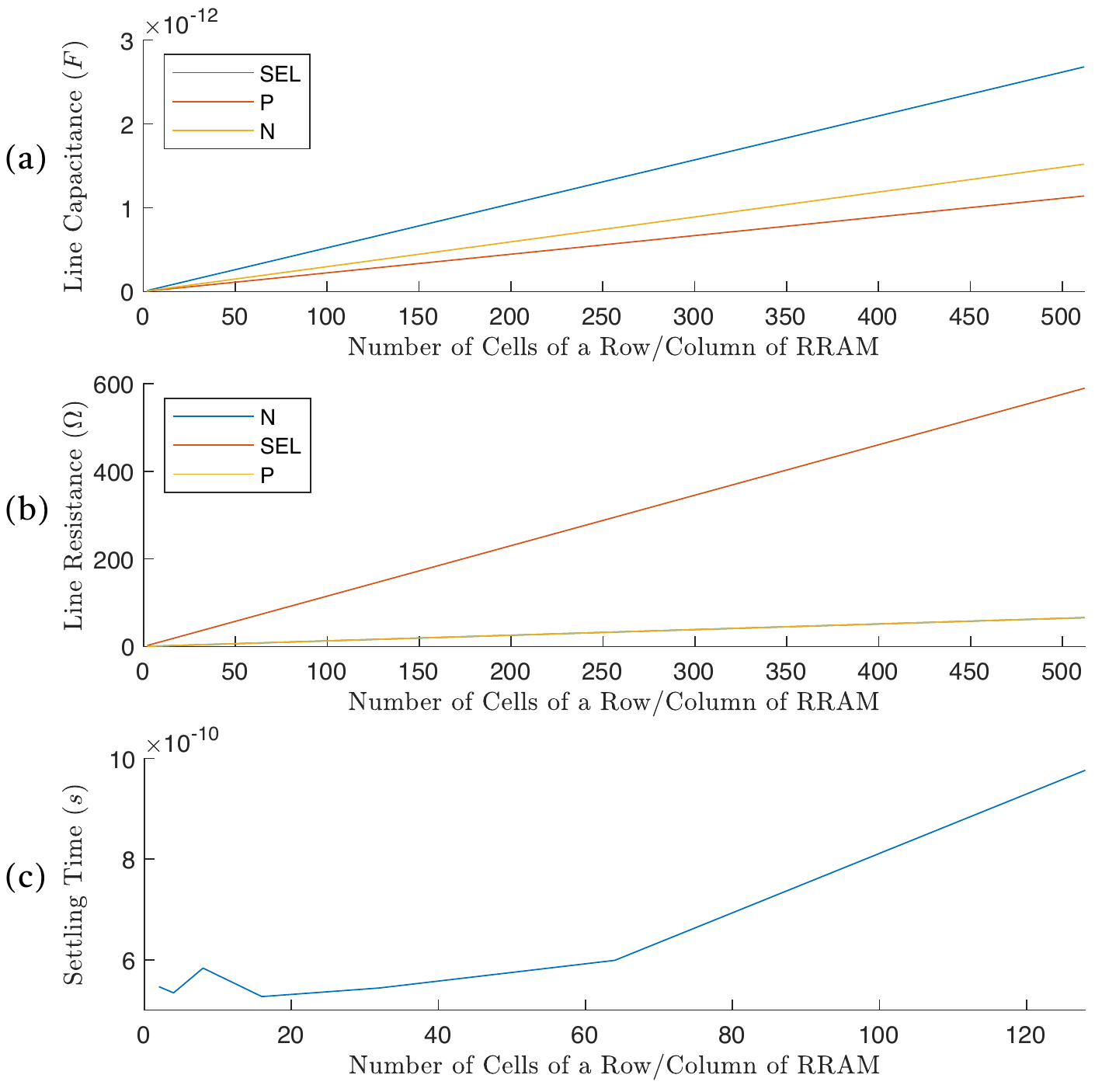}
\caption{(a) Worst case Capacitance and Cross Capacitance ($C+CC$) of lines SEL, P, N with respect to the number of Row/Column cells of RRAM. The increase rates are 5.83\,fF/cell, 3.31\,fF/cell and 2.48\,fF/cell for SEL, N and P lines respectively. (b) Total calculated resistance of SEL, N, P lines with respect to the number of cells of the line. The increase rate is 1.28\,\textOmega/cell for SEL line and 0.14\,\textOmega/cell for N and P lines. (c) Worst case settling time of voltage across the memristor of a cell. This case is for $V_N=$ \emph{VDD}, $V_P=$ \emph{GND} and SS model library. The curve fits the exponential curve $5.223\times 10^{-10}exp(0.004207x)$.}
\label{fig:RRAM_results}
\vspace{-5mm}
\end{figure}

Using the compiler, a number of $N$ x $N$ square RRAM arrays were created. By importing the output files of the compiler in MATLAB, the worst case Capacitances and Cross Capacitances ($C+CC$) were extracted. The results are shown on Fig.~\ref{fig:RRAM_results}a. Fig.~\ref{fig:RRAM_results}a shows that $C+CC$ rises linearly with respect to the number of cells of a line. The SEL lines have greater capacitance. The memory cell has greater horizontal length compared to its vertical width with resemblance to the memory cell layout of Fig.~\ref{fig:RRAM_array_layout}. Lines N and P are shared vertically and occupy less area, thus they suffer of less capacitance. It is worth mentioning that the SEL line capacitance does not include the additional capacitance of the gates of the transistor switches and also that the additional capacitance due to the connections of the cell to the memristor has not been taken into account.

Additionally, the worst case scenario settling time of voltage across the memristor on the post layout views of RRAM cells was investigated. This case was found for \emph{VDD} on N pin and \emph{VSS} on P pin. As it is expected, the worst case was found for the maximum high resistive state where R=5\,M\textOmega. Based on these conditions a testbench was set in Virtuoso and a corner analysis was conducted for a number of RRAM arrays. The settling time range was set equal to 1$\%$ of the final value. Slowest settling time was achieved for Slow NMOS, Slow PMOS (SS) model library. Results are shown on Fig.~\ref{fig:RRAM_results}c. For small array row/column size the settling time is almost stable around 550\,ps, however as the size grows the settling time rises exponentially. This is because both the resistance and the capacitance of the lines increase and therefore the loading time of the line increases.

Based on the SPICE model of the technology library for layout layers, the resistance of lines SEL, N and P were calculated. N and P lines have the same layout shape, thus it was found that they have the same resistance. The results are shown on Fig.~\ref{fig:RRAM_results}b. The resistance also rises linearly with respect to the number of cells of the line. 

The RRAM cell used for the results produces arrays of density equal to 0.082\,Mb/mm\textsuperscript{2} for a 180\,nm technology. The RRAM cells have approximately the same size with SRAM implementations at 180\,nm technology. Assuming linear scaling, the RRAM cells follow the scaling trend of SRAM implementations shown in the below table.
\vspace{-4mm}

\begin{table}[htb]
\caption{Memories density comparison.}
\label{tab:comparison}
\vspace{-2mm}
\begin{center}
\begin{tabular}{c|ccc}
\hline
Ref.                           & Feature Size & Technology & Mb/mm\textsuperscript{2} \\ \hline
\cite{miyano2013highly}        & $40\,$nm         & CMOS       & $0.94$     \\
\cite{toh2011characterization} & $45\,$nm         & CMOS       & $0.33$      \\ 
\cite{guthaus2016openram}      & $45\,$nm         & FreePDK45  & $0.826$     \\
\cite{kushida20090}            & $65\,$nm         & CMOS       & $0.77$      \\
RRAM Compiler $^1$                 & $180\,$nm       & CMOS       & $0.082$      \\ 
\cite{7538307}$^2$                 & $180\,$nm       & CMOS       & $0.067$      \\ 
\cite{guthaus2016openram}      & $0.5\,$\textmu m        & SCMOS  & $0.005$     \\ \hline
\end{tabular}
\end{center}
\raggedright\footnotesize{$^1$ Memory cell array density. $^2$ Memory cell size.}\\
\vspace{-5mm}
\end{table}

\section{Conclusion} \label{sec:conlusions}

To the best of our knowledge, this paper has reported the first open-source memory compiler specifically for RRAM with supporting RRAM architecture. An early version of the compiler has been designed and already been made publically available. Even though, the compiler is a preliminary version, it is able to generate large sized arrays and verify their layout. The authors of this document are currently investigating novel suitable circuits to sense the stored value of a memory cell without altering its resistive state. Next steps include the research and integration in the compiler of write circuits, decoders and multiplexers. 

Besides providing a powerful tool to academia and industry for boosting memory design time, the authors believe that the compiler will facilitate the research of NVMs and RRAMs and their operation.

\section*{Acknowledgement}
The authors would like to thank Yihan Pan for providing the 1T1R cell for use in this memory compiler. This work was in part supported by the Engineering and Physical Sciences Research Council (EPSRC) Programme under Grant EP/R024642/1.

\bibliographystyle{IEEEtran}
\bibliography{IEEEabrv,Section/references}

\end{document}